\documentclass{SciPost}

\hypersetup{
    colorlinks,
    linkcolor={red!50!black},
    citecolor={blue!50!black},
    urlcolor={blue!80!black}
}

\usepackage[bitstream-charter]{mathdesign}
\DeclareSymbolFont{usualmathcal}{OMS}{cmsy}{m}{n}
\DeclareSymbolFontAlphabet{\mathcal}{usualmathcal}

\fancypagestyle{SPstyle}{
\fancyhf{}
\lhead{\colorbox{scipostdeepblue}{\bf \color{white} ~SciPost Physics Proceedings }}
\rhead{{\bf \color{scipostdeepblue} ~Submission }}

\fancyfoot[C]{\textbf{\thepage}}
}

\usepackage{graphicx}
\usepackage{caption}
\usepackage{amsmath}           
\usepackage{lineno} 

\begin{document}

\pagestyle{SPstyle}

\begin{center}{\Large \textbf{\color{scipostdeepblue}{
Using AI on FPGAs for the CMS Overlap Muon Track Finder for the HL-LHC
}}}\end{center}

\begin{center}\textbf{
Pelayo Leguina\textsuperscript{1$\star$},
Santiago Folgueras\textsuperscript{1},
Andrea Cardini\textsuperscript{1},
Elena Aller\textsuperscript{1}
on behalf of the CMS Collaboration
}\end{center}

\begin{center}
{\bf 1} Universidad de Oviedo, Spain
\\[\baselineskip]
$\star$ \href{mailto:leguinapelayo@uniovi.es}{\small leguinapelayo@uniovi.es}
\end{center}

\definecolor{palegray}{gray}{0.95}
\begin{center}
\colorbox{palegray}{
  \begin{tabular}{rr}
  \begin{minipage}{0.37\textwidth}
    \includegraphics[width=60mm]{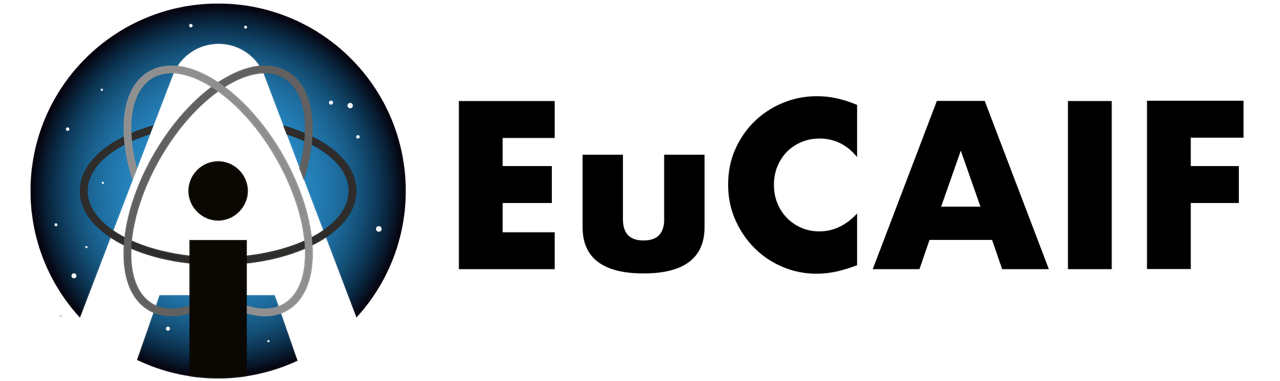}
  \end{minipage}
  &
  \begin{minipage}{0.5\textwidth}
    \vspace{5pt}
    \vspace{0.5\baselineskip} 
    \begin{center} \hspace{5pt}
    {\it The 2nd European AI for Fundamental \\Physics Conference (EuCAIFCon2025)} \\
    {\it Cagliari, Sardinia, 16-20 June 2025
    }
    \vspace{0.5\baselineskip} 
    \vspace{5pt}
    \end{center}
    
  \end{minipage}
\end{tabular}
}
\end{center}

\section*{\color{scipostdeepblue}{Abstract}}
\textbf{\boldmath{%
Operating the CMS Level-1 trigger under the intense conditions of the High-Luminosity Large Hadron Collider—with approximately 63~Tb/s of input and a fixed 12.5~$\mu$s latency—poses a demanding real-time reconstruction challenge. The CMS muon system is organized into three regions: a barrel, an endcap, and the intermediate barrel–endcap ``overlap'' region. In this overlap transition, the Overlap Muon Track Finder can be suboptimal for displaced-muon and long-lived-particle signatures. We present a first approach to a graph neural network tailored to these constraints, using GraphSAGE layers and a compact multi-layer perceptron to regress the inverse transverse momentum of muons. A PyTorch to C\texttt{++} and high-level synthesis flow demonstrates feasibility, with initial results showing good agreement with simulation. Although a fully parallel implementation would exceed available field-programmable gate array resources, quantization, pruning, and multiplier reuse point the way toward a practical Phase-2 deployment.
}}

\vspace{\baselineskip}

\noindent\textcolor{white!90!black}{%
\fbox{\parbox{0.975\linewidth}{%
\textcolor{white!40!black}{\begin{tabular}{lr}%
  \begin{minipage}{0.6\textwidth}%
    {\small Copyright attribution to authors. \newline
    This work is a submission to SciPost Phys. Proc. \newline
    License information to appear upon publication. \newline
    Publication information to appear upon publication.}
  \end{minipage} & \begin{minipage}{0.4\textwidth}
    {\small Received Date \newline Accepted Date \newline Published Date}%
  \end{minipage}
\end{tabular}}
}}
}


\section{Introduction}
\label{sec:intro}
\begin{figure}[h]
\centering
\includegraphics[width=0.8\linewidth]{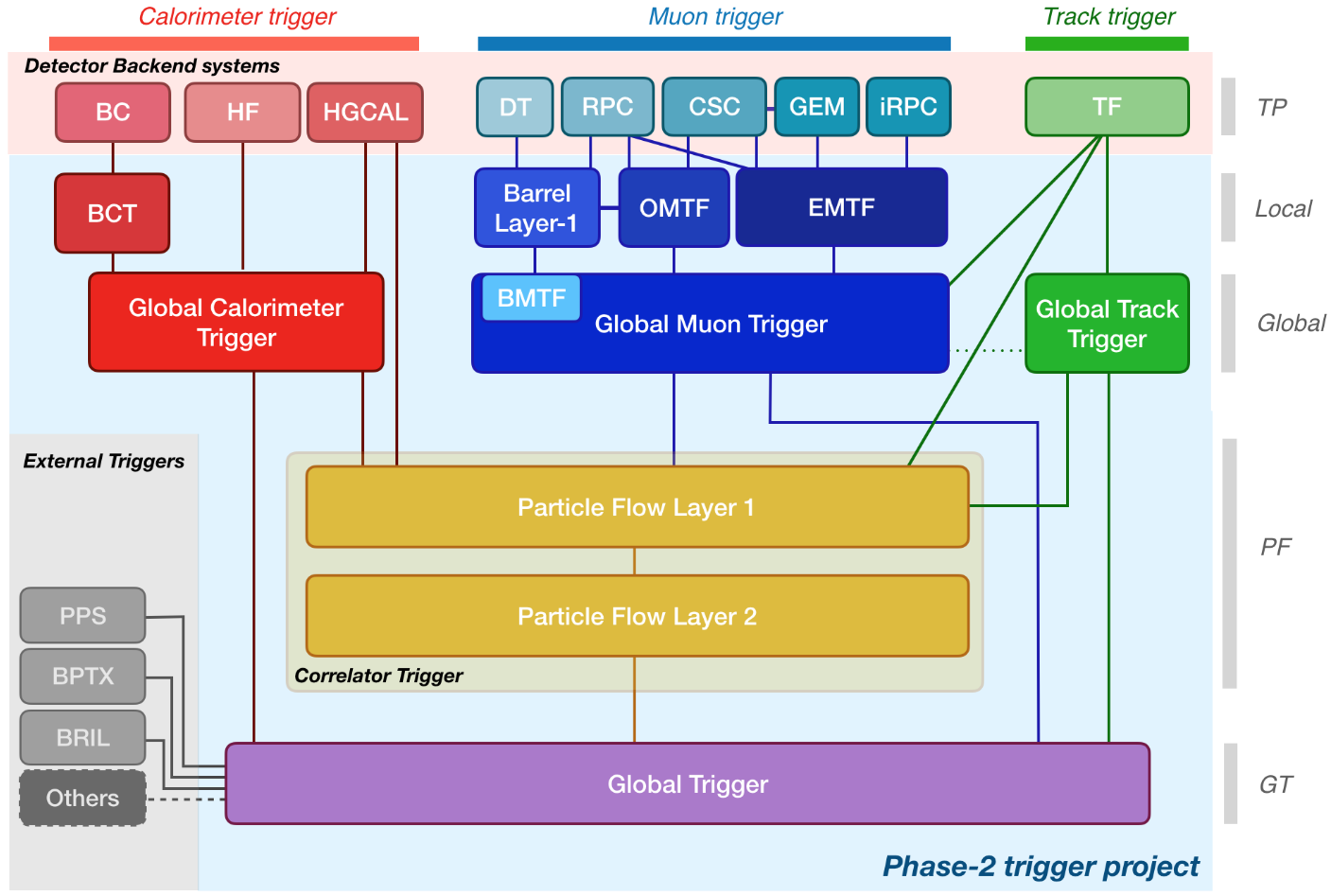}
\caption{Functional diagram of the CMS Level-1 Phase-2 upgraded trigger design~\cite{CMS_PHASE2_TDR}.}
\label{fig:l1t}
\end{figure}

The CMS experiment~\cite{CMS_JINST} at the CERN LHC features a trigger system designed primarily to select interesting physics processes from the 40~MHz bunch crossing rate and, as a consequence, reduce the data rate to manageable levels.
The CMS muon system, described in the technical design report~\cite{CMS_MUON_TDR}, provides coverage with Drift Tubes (DT), Cathode Strip Chambers (CSC), and Resistive Plate Chambers (RPC).
The CMS Level-1 Trigger (L1T)~\cite{CMS_TRIGGER} must process data under tight latency constraints.
During Run~2, the maximum latency was 3.8~$\mu$s, and it was increased to 12.5~$\mu$s for the Phase-2 upgrade to accommodate inputs from the upgraded silicon tracker track trigger and the high-granularity calorimeter~\cite{CMS_PHASE2_TDR}.
The new architecture features high-bandwidth optical links up to 28~Gb/s and state-of-the-art Xilinx VU13P FPGAs, opening the possibility of exploiting advanced machine learning models in hardware, while respecting the fixed-latency budget.
Figure~\ref{fig:l1t} summarizes the Phase-2 L1 trigger and motivates FPGA-native AI algorithms within this envelope.

\section{The Overlap Muon Track Finder}
The Overlap Muon Track Finder (OMTF) reconstructs muons in the barrel–endcap transition, combining information from DT, RPC, and CSC chambers~\cite{OMTF_JINST}.
The current algorithm compares detected stubs (short local track segments built within a single station) to reference patterns of muon tracks at given transverse momenta.
This approach has been effective for prompt muons but is less efficient for muons originating from displaced vertices or for signatures associated with long-lived particles.
A graph representation of the OMTF region is particularly natural: nodes correspond to stubs measured in the DT and CSC stations (and associated RPC planes), and edges encode local geometric relationships via $(\Delta\eta,\Delta\phi)$ between stubs.
This preserves sparsity and the irregular geometry of the overlap region and provides a suitable substrate for graph neural networks.

\section{Graph Neural Network architecture}
\begin{figure}[h]
\centering
\includegraphics[width=\linewidth]{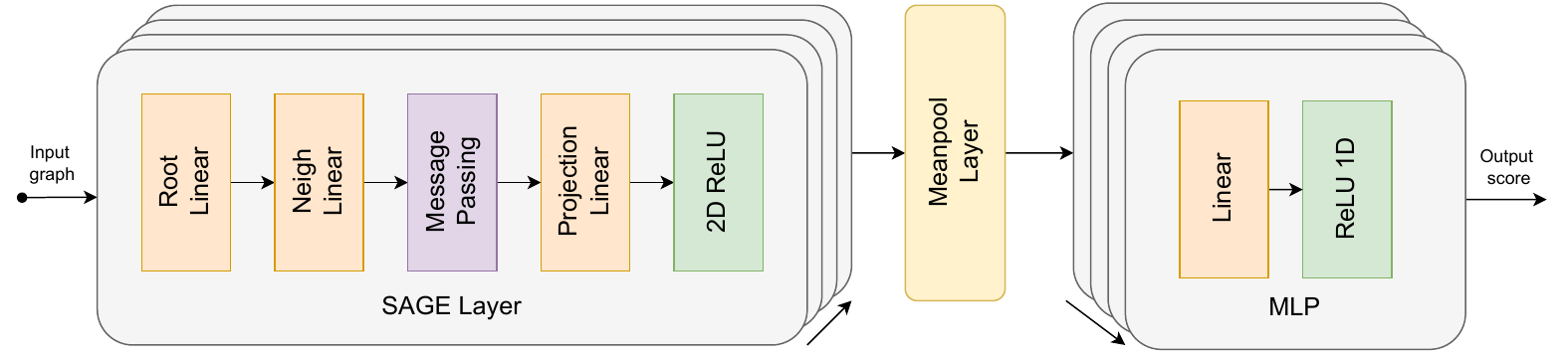}
\caption{Block diagram of the GNN architecture implemented in HLS: four SAGE convolution blocks (each internally composed of three Linear transforms, message passing, and ReLU-2D), a global mean pooling layer, and a four-layer MLP head for $q/p_T$ regression.}
\label{fig:arch}
\end{figure}

Figure~\ref{fig:arch} shows the model architecture, which consists of four GraphSAGE convolutional layers~\cite{arXiv:1706.02216}, followed by a global mean pooling stage and a four-layer MLP head.
Each GraphSAGE layer updates node embeddings by combining a learned projection of the node’s own features with an aggregated projection of the features of its neighbors.
In our implementation, each SAGE block is composed of three Linear transforms, a message-passing aggregation over the edge list, and a two-dimensional ReLU applied to the per-node feature maps; an optional $\ell_2$ normalization can be enabled after combination.
After the four SAGE layers, global mean pooling reduces the variable-size graph to a fixed-dimension vector that is processed by a four-layer MLP to produce the final scalar prediction $q/p_T$.
In total, the sequence is: SAGEConv ($d_{\text{in}}\!\to\!4h$), SAGEConv ($4h\!\to\!2h$), SAGEConv ($2h\!\to\!2h$), SAGEConv ($2h\!\to\!2h$), global mean pool, and four Linear layers that progressively reduce the hidden dimension to one output.

\medskip
\noindent\emph{Why this architecture for the OMTF physics problem?}
The overlap region is sparse and geometrically irregular; event activity varies from a handful of stubs to dense topologies.
SAGE-style neighborhood aggregation captures local correlations among nearby stations (e.g.\ DT–RPC–CSC consistency) while remaining robust to missing inputs, and the global pooling + MLP summarize global track-level cues (overall curvature, sign) in a fixed-latency mapping.
This local-to-global design allows efficient handling of variable multiplicity, is tolerant to noise, and naturally extends to displaced signatures where muon trajectories deviate from prompt-track templates.

\section{Feasibility of the implementation into FPGAs}
Each block in Fig.~\ref{fig:arch} is mapped to a modular C++ implementation synthesized with HLS: the SAGE convolution realizes the three Linear projections, message passing over the compact edge list, feature combination, and activation; Linear layers implement dense matrix multiplications; the mean pooling layer reduces per-node embeddings to a graph-level vector; and lightweight stages provide normalization and activations where needed.
To ensure functional correctness, each layer has been independently implemented in HLS and tested against the PyTorch reference.
This was achieved by modularizing the layer definitions in Python, extracting their main operations, re-coding them in C++, and adapting the interfaces for HLS.
This approach preserves the behaviour of the PyTorch model while progressively building verified hardware components.

The computational load of the model can be expressed as
\begin{equation}
\mathcal{O}\!\left(\sum_{\ell=1}^{L} 
\Big[ |V|\, d_{\text{in}}^{(\ell)} d_{\text{out}}^{(\ell)} 
+ |E|\, d_{\text{out}}^{(\ell)} \Big]\right) ,
\label{eq:comp-load}
\end{equation}
where $|V|$ and $|E|$ are the number of nodes and edges in the graph, and 
$d_{\text{in}}^{(\ell)}, d_{\text{out}}^{(\ell)}$ are the input and output dimensions of layer $\ell$.
The first term accounts for node-wise linear transformations, while the second corresponds to neighbor messages sent along edges.

In the OMTF context, the graph is constructed from the ten detector layers that provide the most relevant stub information for muon track matching across DT, CSC, and RPC chambers, which motivates $|V|=10$.
The number of possible edges corresponds to the pairwise connections between nodes, bounded by ${10 \choose 2}=45$, hence $|E|\leq 45$.

For a representative configuration with hidden dimensions 
[3$\to$128], [128$\to$64], [64$\to$64], [64$\to$64] for the four SAGE convolutions, followed by four linear layers up to $64\times64$, the total multiply–accumulate count of the GraphSAGE part alone is approximately
\begin{equation}
\sum_{\ell=1}^{4} \mathrm{MACs}_{\text{Conv}\,\ell} \;\approx\; 335{,}360 .
\end{equation}

\noindent\textbf{Target device and resources.}
Our target is a Xilinx Virtex UltraScale+ VU13P (XCVU13P), which provides on the order of $10^4$ DSP48E2 slices and substantial on-chip BRAM/URAM capacity.
A naively unrolled, fully parallel implementation of the above model in 16-bit fixed-point arithmetic would exceed the available DSP budget, motivating architecture-level optimizations.

\medskip
\noindent\textbf{Optimization strategy.}
We consider four complementary tactics:
(i) \emph{Quantization}: 8-bit (and exploratory 4-bit) weights/activations with quantization-aware training to preserve regression fidelity;
(ii) \emph{Pruning}: structured sparsification of edges and hidden channels guided by physics performance to reduce MACs and memory bandwidth;
(iii) \emph{Multiplier reuse}: reuse-factor sweeps to time-multiplex multipliers while maintaining an initiation interval compatible with the L1T envelope;
(iv) \emph{Model simplification}: reduction of hidden dimensions and, where justified by ablation studies, a decrease in the number of hidden layers to lower computational cost and resource usage while preserving physics performance.
For small dense layers, hybrid DSP/LUT multipliers can reduce pressure on DSP resources.
Combined, these measures aim to reach a feasible point on VU13P within fixed-latency constraints.

\section{Further developments}
To meet L1T throughput and latency targets, we are extending the optimization program above and integrating fixed-latency scheduling across modules.
Quantization to 8-bit (and selectively 4-bit) is being validated with QAT; edge/channel pruning exploits detector sparsity; reuse-factor sweeps share multipliers without violating the per-graph initiation interval.
In addition, we explore partitioning strategies across BRAM and the use of LUT-based multipliers in peripheral layers.
Together, these measures are expected to bring the implementation within the available resources of Phase-2 FPGA platforms.

\section{Conclusion}
We have presented a first proof-of-concept of a GNN-based muon reconstruction for the CMS Overlap Muon Track Finder with a strong emphasis on its hardware realization on FPGAs.
The modular architecture, developed in Python and systematically translated into C++ and HLS, allowed us to reproduce the behavior of the PyTorch model layer by layer and validate the design at each stage of the hardware flow.
While a fully parallel mapping exceeds the resource capacity of the Xilinx VU13P, the study demonstrates that with the proposed optimizations the design can fit within the constraints of the CMS Level-1 Trigger Phase-2 upgrade.
This work highlights the feasibility of deploying AI-based reconstruction in real-time trigger systems, underlining the importance of hardware-oriented design choices in high-energy physics environments.

\section*{Acknowledgements}

This work and the authors are fully supported by ERC grant (INTREPID, 101115353).   
Funded by the European Union. 

\bibliographystyle{SciPost_bibstyle}

\end{document}